%% file: main.tex
\documentclass{acm_proc_article-sp}

\usepackage{rotating, booktabs, enumitem, amsmath, balance}
\usepackage[hyphens,obeyspaces,spaces]{url}
\usepackage{hyperref}
\hypersetup{
    pdftitle={CrowdSTAR: A Social Task Routing Framework for Online Communities},    
    pdfauthor={Besmira Nushi, Omar Alonso, Martin Hentschel, and Vasileios Kandylas},     
    pdfsubject={CrowdSTAR},   
    pdfcreator={Besmira Nushi},   
    pdfproducer={},  
    pdfkeywords={Task Routing, Social Search, Question Answering, Crowdsourcing, User Expertise}, 
    hidelinks
 }
\usepackage{tikz}
\usetikzlibrary{patterns}
\usetikzlibrary{shapes.multipart}
\usetikzlibrary{automata,positioning}
\usepackage{pgfplots}
\usepgflibrary{patterns}
\usetikzlibrary{patterns}
\newenvironment{my_enumerate}{
   \begin{enumerate}
   \setlength{\itemsep}{1pt}
   \setlength{\parskip}{0pt}
   \setlength{\parsep}{0pt}
}{
   \end{enumerate}
}
\newtheorem{example}{Example}
\DeclareGraphicsExtensions{.pdf,.png,.jpg}
\pgfdeclareimage{user}{user}
\usetikzlibrary{fit,positioning,shapes,arrows,decorations.pathreplacing,intersections}

\begin{document}

\title{CrowdSTAR: A Social Task Routing Framework for Online Communities}

\numberofauthors{4} 
\author{
\alignauthor
Besmira Nushi\\
       \affaddr{ETH Zurich}\\
       \affaddr{Systems Group}\\
       \email{nushib@inf.ethz.ch}
\alignauthor
Omar Alonso\\
       \affaddr{Microsoft}\\
       \affaddr{Silicon Valley}\\
       \email{omar.alonso@microsoft.com}
\and       
\alignauthor
Martin Hentschel\\
       \affaddr{Microsoft}\\
       \affaddr{Silicon Valley}\\
       \email{hemartin@microsoft.com}
\alignauthor
Vasileios Kandylas\\
       \affaddr{Microsoft}\\
       \affaddr{Silicon Valley}\\
       \email{vakandyl@microsoft.com}
}
\maketitle
\begin{abstract}
The online communities available on the Web have shown to be significantly interactive and capable of collectively solving difficult tasks. Nevertheless, it is still a challenge to decide how a task should be dispatched through the network due to the high diversity of the communities and the dynamically changing expertise and social availability of their members. We introduce CrowdSTAR, a framework designed to route tasks across and within online crowds. CrowdSTAR indexes the topic-specific expertise and social features of the crowd contributors and then uses a routing algorithm, which suggests the best sources to ask based on the knowledge vs. availability trade-offs. We experimented with the proposed framework for question and answering scenarios by using two popular social networks as crowd candidates: Twitter and Quora.
\end{abstract}




\input{sections/sec1_introduction}
\input{sections/sec2_relatedwork}
\input{sections/sec3_humansvsmachines}
\input{sections/sec4_userutilitymodel}
\input{sections/sec5_socialrouting}
\input{sections/sec6_crowdstar}
\input{sections/sec7_experiments}
\input{sections/sec8_conclusions}

\balance
\small
\bibliographystyle{abbrv}
\bibliography{sigproc}

\end{document}

%% file: sections/sec1_introduction.tex
\section{Introduction}

Crowdsourcing emerged as a methodology to solve problems whose solutions require human intervention and skills and
therefore cannot be solved by machines only. There are several crowdsourcing platforms available, each one with its own
characteristics and niche market. Nonetheless, the networks of people in these platforms are not yet interconnected and
the distributed wisdom of crowds and their respective users is not catalogued. Because of this rich crowdsourcing
environment it is difficult to say which of these sources is most appropriate to solve a given problem. The challenge we
are addressing in this paper is how to find the best experts within the best matching crowd for a given task. This challenge is made more
difficult by the fact that the characteristics of networks and users change dynamically over time. For example, the user
base of a crowdsourcing platform or social network might significantly grow or shrink over time. The contribution of a
single user may vary from being absolutely committed to only being marginally present or not present at all. A solution
that finds best-matching crowds therefore has to continuously adapt to such a dynamic setting.

We believe that the potential of the collective wisdom of online crowdsourcing can be fully explored only if there exist services that perform the following functions:
\begin{enumerate}
\item Detecting and indexing the dynamically changing crowd expertise.
\item Using this knowledge index to route tasks to the right crowds and individual users.
\end{enumerate}
Indexing the knowledge and availability characteristics of a crowd and its users allows us to route tasks to the right crowd and users. We call this approach \emph{social task routing}. The intent of social task routing is to find competent people that answer a problem for which an otherwise definite answer does not exist (e.g., via Web search).  The following example is one of such problems:
\begin{example}
Alice is researching the impact of academia on startup companies in the Bay Area. In particular, she is interested in a list of professors from well-known universities in the Bay Area who founded a startup.
\end{example}
Alice tries to input various queries into search engines but none of her formulations are giving her a satisfying
answer. The specific answer to this problem is nowhere aggregated on the Web but needs to be collected from people who
are knowledgeable in the field of startups and academia. The goal of social task routing in this case is to find out
which of the online communities contain this expertise and which individuals in the community to ask.

The main contribution of this paper is a system for social task routing that combines expertise detection with interactivity and availability
characteristics of users. As we will describe in the following sections, expertise detection is a crucial factor for the
accuracy of a social task router, yet it is not sufficient. Equally important factors are
the users' interactivity and availability characteristics. Therefore, the second main contribution of this paper 
is an exploration of the trade-offs between these dimensions. Our system, \textbf{CrowdSTAR}
(\textbf{Crowd}sourced \textbf{S}ocial \textbf{Ta}sk \textbf{R}outing), investigates these aspects in the context of
question-answering tasks using two popular social networks, Quora and Twitter.

The rest of this paper is organized as follows. We first cover related work in Section~\ref{sec:relatedwork}. We then present the results of a
survey on the usability of Q\&A systems in Section~\ref{sec:humansvsmachines}. We define the user utility model in Section~\ref{sec:utilitymodel} that is used to
support the social task routing algorithm described in Section~\ref{sec:taskrouting}. We explain the system architecture of CrowdSTAR in Section~\ref{sec:crowdstar}
and present the results of the evaluation of our system in Section~\ref{sec:experiments}. Section~\ref{sec:concl} concludes this paper.

%% file: sections/sec2_relatedwork.tex
\section{Related Work}
\label{sec:relatedwork}

Social task routing has been analyzed in several studies and stands at the boundary of major research fields like
collaborative information seeking and crowdsourcing. Dustdar and Gaedke~\cite{dustdar2011social} were among the first to
envision the general social routing principle supported by Web-scale workflows. Law and von Ahn~\cite{von2009human}
provide insights into the different forms of task routing and recommendation for crowdsourcing. Morris, Teevan, and
Panovich~\cite{morris2010comparison} describe a thorough comparison between Web search and social search (i.e.
forwarding the question to a social network). Their work confirms that none of the searching methods is a general
panacea but that they work best if combined together. Further studies by Jeong et al.~\cite{jeong2013crowd} and Morris
et al.~\cite{morris2010people} in the same context show that answers from humans increase the users' satisfaction and
confidence. Indeed, Oeldorf-Hirsch et al.~\cite{oeldorf2014search} examined that, when given the option of using a
search engine and/or a social network, users use status message question asking for 20\% of their information needs.
Ellison et al.~\cite{ellison2013calling} study requests for responses on Facebook (questions and other forms of
requests) and categorize those by cost and type.

The most relevant work to our problem definition is the one presented by Bozzon et al.~\cite{bozzon2013choosing}. The
authors propose a general resource-to-user graph model to represent any given crowd. They then compute user scores based
on the frequency of the search term within the resources associated to the user or their graph neighbors. The method
though does not take into account the social features of the experts (interactivity and availability) and bases the
crowd selection on expertise only. Further analyses focus on the Q\&A potential of single crowds but do not make use of
the social features \cite{zhou2009routing,li2011question}. Furthermore, Mamykina et al. \cite{mamykina2011design} and
Paul, Hong, and Chi~\cite{paul2012authoritative} study
Q\&A communities and their respective incentives and reward strategies concluding that building a good online reputation
is one of the most important motivations.

Horowitz and Kamvar~\cite{horowitz2010anatomy} explore the concept of social availability. Their work
characterizes a social search engine where people ask questions to other users via different means of communication, for
example, email and instant messaging. The availability of the members is incorporated by applying filters that prevent
asking people who are not online at the moment or have been asked very recently. Thus, it is not part of the user-topic
model but works as a general pruning criterion. The expert search is isolated within a single network and within the
circle of contacts of the person who is asking the question. As we show in the next section, although this can be
efficient for personal questions, it might not be as profitable for questions requiring a broader domain of competence.
Sung, Lee, and Lee \cite{sung2013booming} introduce a model that linearly combines (topical) availability
with expertise into a single measure called question affordance.

Gathering expertise evidence in social networks is an active field of research
\cite{balog2012expertise,pal2011identifying,bozzon2013choosing,ghosh2012cognos}. The generalized approach is to rank
candidate experts according to the likelihood of person $e$ being an expert on query $q$
\cite{balog2012expertise}. This likelihood is typically computed by combining different features of the candidate expert
given the query $q$. Pal and Counts define multiple features around the textual content a person generates (e.g., tweets
on topic, mention impact) \cite{pal2011identifying}. Bozzon et al.~\cite{bozzon2013choosing} use term frequency and
inverse document frequency (tf/idf) of words and entities as discriminating features. Ghosh et
al.~\cite{ghosh2012cognos} use the number of appearances of a person in Twitter lists to compute the likelihood of the
person being an expert. These approaches provide good results but differ in detail for individual queries. The user
metrics explained in this study are inspired by Pal and
Counts~\cite{pal2011identifying} and elaborated for a better depiction of our vision.

%% file: sections/sec3_humansvsmachines.tex
\section{Humans vs. Machines}
\label{sec:humansvsmachines}

Q\&A communities are an effective mechanism for information seeking on the web. They contain a large number of questions
with answers which motivates a user to try a Q\&A service instead of searching for web pages on a search engine. For a
deeper understanding of the user requirements, it is interesting to know which types of questions a user would likely
ask to another user or try to find the answer using a search engine. To quantify these preferences, we designed an
experiment that consisted of showing to a crowdsourcing worker an existing question taken from Quora (with no user interface treatment) and asking two survey
questions about it:
\begin{enumerate}
  \item Do you think that you can find the answer to this question by typing a few keywords on a search engine or would
  you prefer a direct answer by a person (a friend or expert)?
  \item If you prefer a human answer for this particular question, would you further prefer a friend or an expert providing the answer?
\end{enumerate}

\begin{table}[t]
\centering
\caption{Distribution of most frequent interrogative words of the top 500 and bottom 500 questions on Quora.}
\label{tab:heads}
\begin{tabular}{@{}lcc@{}}
  \toprule
  Interrogative word      & Top 500 	& Bottom 500	\\ \midrule
  What      & 60.8\% 	& 31.2\%		\\ 
  How       & 7.6\% 	& 17.6\%		\\
  Which     & 6.8\% 	& 2.4\%			\\
  Why       & 4.8\% 	& 6.8\%			\\
  Who       & 4.4\% 	& 1.6\%			\\ 
  Where     & 1.2\% 	& 3.8\%			\\ \bottomrule
\end{tabular}
\end{table}

\begin{table}[t]
\centering
\caption{Results of Question 1: ``Do you think you can find the answer to this question using a search engine or through a person?''}
\label{tab:question1}
\begin{tabular}{@{}lcc@{}}
  \toprule
  Answer via     & Top 500 & Bottom 500	\\ \midrule
  Search engine  & 53.8\%  & 75.1\%		\\
  Person         & 46.2\%  & 24.9\%		\\ \bottomrule
\end{tabular}
\end{table}

\begin{table}[ht!]
\centering
\caption{Results of Question 2: ``Would you prefer a friend or an expert providing the answer?''}
\label{tab:question2}
\begin{tabular}{@{}lcc@{}}
  \toprule
  Answer via 		& Top 500 	& Bottom 500	\\ \midrule
  Expert 			& 56.2\% 	& 60.8\%		\\ 
  Friend     		& 22.9\% 	& 17.8\%		\\ 
  No preference     & 20.9\% 	& 21.4\%		\\ \bottomrule
\end{tabular}
\end{table}

The sample data set consisted of the top 500
questions (with the most number of answers) and the bottom 500 questions (with the least number of answers) from Quora.
Some examples of the questions we asked are: Why is U2 so popular? or What are the best storytelling
songs? We asked 3 judges per question and took the majority vote as the final answer.

To differentiate the two sets of questions asked, we group the questions by interrogative words, or question words (for
example \emph{what}, \emph{when}, and \emph{how}). For the top 500 questions there are 28 unique interrogative words,
whereas for the bottom 500 questions there are 47 unique interrogative words. Table~\ref{tab:heads} shows the
distribution of the most frequent interrogative words of the two data sets. The interrogative pronoun \emph{what}
dominates the top 500 questions, while it is only half as frequent in the bottom 500 questions. On the other hand, the
interrogative \emph{how} is much more frequent in the bottom 500 questions than in the top 500. Our findings are that
the top 500 questions are more \emph{advice-type} questions (e.g., What do you do when no one believes in you?) or
questions that require perspective (e.g., Can money buy happiness?), whereas the bottom 500 questions are more
\emph{fact-finding} questions (e.g., What is knowledge management? and How much did Facebook acquire Beluga for?).

The results of Question 1, ``Do you think you can find the answer to this question using a search engine or through a
person?'', are shown in Table~\ref{tab:question1}. For the top 500 questions, the judges slightly prefer to find an 
answer using a search engine than by asking a person. For the bottom 500 questions, a search engine is preferred in 75\%
of the cases. These differences can be explained by the different types of questions in the two data sets. The top 500
questions more frequently ask for advice or require perspective and thus can be better answered by asking a person. The
bottom 500 questions more frequently ask for facts and thus can be better answered using a search engine.

The results of Question 2, ``Would you prefer a friend or an expert providing the answer?'', are summarized in
Table~\ref{tab:question2}. Overall, answers by experts are preferred over answers by friends by a significant
difference for both data sets. For the bottom 500 questions, the preference for answers by friends drops compared to the
top 500 questions while the preference for experts increases. Our explanation is that for questions that require
\emph{how to}, users trust more an answer from an expert than from a friend. The preference for experts justifies
the need for expertise detection before asking questions in Q\&A communities. Another observation is that questions
preferred to be answered by humans contain many non-factual/non-factoid answers. Questions starting with \emph{why} and
\emph{what was the reason} show a preference to be answered by humans. In contrast, those starting with \emph{what is the
name}, \emph{who}, and \emph{how long} can generally be satisfied by a search engine.


%% file: sections/sec4_userutilitymodel.tex
\section{User Utility Model}
\label{sec:utilitymodel}

In this section we describe the user model from a utility perspective. We emphasize the fact that the utility of a crowd
member (i.e., her adequacy to solve a given task) is (1) topic-specific, (2) continuously changing, and (3) strongly
affected by the user's social behavior in the network. In contrast to previous work, we decide to model more
than one feature for each triple $<user, topic, crowd>$ and use them altogether in the routing technique that we propose
later on.

We first identify two main dimensions for a given user part of a certain crowd on a particular topic: \emph{Knowledge}
and \emph{Availability}. \emph{Knowledge} is the dimension that captures the passive or active expertise on the topic
while \emph{Availability} shows the social involvement in answering questions or conversing on the same topic. From our
task routing experiments in Twitter and Quora aiming for high knowledge
is crucial but not enough. Accounts which seem to know a lot on a particular matter can be slow or not helpful in answering
questions. For example, \emph{BBC News} might be a good resource for news retrieval but it is rarely collaborating in
task solving through conversations. On the other hand, a talkative account without expertise may give answers with low
quality.

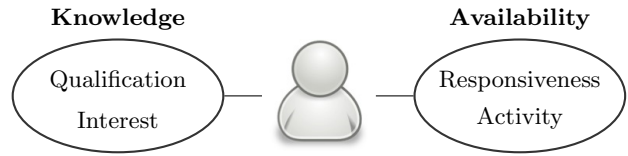
\begin{figure}[t]
  \centering
	\begin{tikzpicture}
		\tikzstyle{texts}=[draw=none,fill=none]
		\tikzstyle{ell}=[ellipse, minimum width = 28mm, minimum height = 15mm, thick, draw=black!80, node distance = 16mm]
		\tikzstyle{line}=[draw]
		\node[inner sep=0pt] (U) at (0, 0){\includegraphics[width=1.5cm]{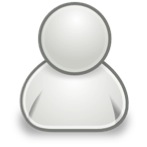}};
		\node[ell] (K) [left=0.5 cm of U, label=above:{\bf Knowledge}] { };
		\node[ell] (R) [right=0.5 cm of U, label=above:{\bf Availability}] { };
		\node[texts] (T1) [below=-0.7 cm of K, label=above:Qualification] { };
		\node[texts] (T2) [below=0.25 cm of T1, label=above:Interest]  { };
		\node[texts] (T3) [below=-0.7 cm of R, label=above:Responsiveness]  { };
		\node[texts] (T4) [below=0.25 cm of T3, label=above:Activity]  { };
		\path [line] (U) -- (K);
		\path [line] (U) -- (R);
	\end{tikzpicture}
	\caption{User utility model for a given topic.}
	\label{fig:UserModel}
\end{figure}

We improve the definition of these two dimensions by decomposing them into two other sub-features. As shown in
Figure~\ref{fig:UserModel}, \emph{Knowledge} is further divided into \emph{Qualification} and \emph{Interest} while
\emph{Availability} is broken down into \emph{Responsiveness} and \emph{Activity}. Semantically, the meaning and the
intent of each of the social availability features is as follows:
\begin{my_enumerate}
\item \emph{Qualification} ($K_1$): How much original and qualitative content does the user generate? A user on Quora, for example, may be active on a subject by posting questions but this does not show that he or she is qualified.
\item \emph{Interest} ($K_2$): How active and interested is the user? The aim is to compute the degree of interest on the topic with respect to the overall activity in the network.
\item \emph{Responsiveness} ($A_1$): How responsive is the user to conversations and questions relevant to the topic? This metric not only can be useful to retrieve answers faster but also can be exploited as a discriminative filter for distinguishing advertisement/company accounts from real human members.
\item \emph{Activity} ($A_2$): How long has it been since the user's last contribution on the topic? Considering that human crowd members cannot be accessed continuously, this metric 
helps to increase user satisfaction by keeping them engaged without overloading. 
\end{my_enumerate}

\subsection{Expertise detection}

Expertise detection is the problem of finding people who are experts. The field of research that studies expertise
detection is called \emph{expertise retrieval}. As shown in Section~\ref{sec:humansvsmachines}, we need experts to help
us answer questions that we are unable to answer using a search engine on the Web. Even with lots of information at
hand, we need experts to classify information, draw conclusions, and present us with an answer. An excellent summary of
the research findings in expertise retrieval is provided in the recent survey by Balog et al.~\cite{balog2012expertise}

\begin{table}[t]
\caption{Metric definitions of expertise.}
\centering
\label{tab:expertise_metrics}
\begin{tabular}{@{}p{2cm} l@{}}
  \toprule
  Metric & Definition \\ \midrule
  A      & answer \\
  CA     & correct answer \\
  P      & post \\
  OP     & original non-conversational post \\ \bottomrule
\end{tabular}
\end{table}

There are two main challenges of expertise detection: \emph{candidate selection} and gathering \emph{expertise evidence}. Candidate
selection is the problem of finding candidate experts on a particular topic. A candidate expert can either be an author
of a text or a person that is referenced in a text, via mention or citation \cite{balog2012expertise}. Gathering
expertise evidence is the problem of determining the strength of expertise of a candidate expert given the textual
evidence. The strength of expertise is the likelihood of the person being an expert, given the text documents,
characteristics of documents, and type of candidate selection. Interestingly, the challenges of expertise detection,
candidate selection and gathering expertise evidence, are comparable to the challenges in Web search, namely document
matching and document ranking.

Candidate selection in social networks is straightforward because authors of texts can be identified by unique
social network identifiers. These identifiers are unique and permanent. Moreover, references to other
persons are typically made via quasi-permanent usernames that hardly ever change. On Twitter, for example, each user has
a unique numeric user id (e.g., \emph{1969163161}) and a
quasi-permanent username (e.g., \emph{@SocialQARouting}). Unique user ids and quasi-permanent
usernames make it feasible to select candidate experts via authorship and references. Candidate selection is hard if
references are real names, part of names, or abbreviated names. For example, if people tweet about \emph{``Social Q}\&\emph{A
Routing''} instead of
referring to \emph{@SocialQARouting}, it is difficult to infer that these mentions refer to the same user on Twitter.
Resolving such cases requires entity resolution techniques \cite{bekkerman2005disambiguating, benjelloun2009swoosh}. In this paper we do not
utilize entity resolution; we select candidate experts by user id and
username only.

Candidate selection in our approach is achieved via two steps: finding user-generated documents and selecting the
authors of these documents as candidate experts. In the first step, we find user-generated documents by matching all
documents of a social network on a particular topic. Matching in Twitter is performed by checking whether the topic is
contained in a document. In Quora, the content is tagged by users or editors with the topics it belongs to. In the
second step, we choose as candidate experts the set of authors of the matched documents. Gathering expertise evidence is
based on the two features of the Knowledge dimension: qualification and interest, whose meaning was introduced earlier
in this section. Formally, qualification ($K_1$) and interest ($K_2$) are defined as follows for a social network crowd
$c$, user $u$, and topic $t$ (the metrics used in the formulas are explained in Table~\ref{tab:expertise_metrics}):
\begin{align*}
K_1(c,u,t) &= \frac{\text{CA}_{(c,u,t)}}{\text{A}_{(c,u,t)}} + \frac{\text{OP}_{(c,u,t)}}{\text{P}_{(c,u,t)}} \\
K_2(c,u,t) &= \frac{\text{P}_{(c,u,t)}}{\text{P}_{(c,u)}}
\end{align*}
On Twitter, a post is a tweet message. An original post is a non-conversational tweet that is not a retweet. A
non-conversational tweet is a tweet that is not addressed to another user and thus is visible to all followers of the
tweeting user. On Quora, a post is an answer, a question, or a blog post that does not include Q\&A. An original post is
an answer or a blog post, but not a question because questions do not provide evidence of a user's qualification. The
definition of correct answer depends on the social network as well. For Quora, we consider an answer correct if it
received at least two \emph{upvotes}. For Twitter, we manually judge answers as correct or incorrect. It must be noted
that manually judging answers does not scale with the number of questions answered. An alternative is to use an
additional crowdsourcing step to let crowd workers upvote answers by Twitter users, which in fact happens by design in
Quora.

The above definition of qualification and interest is susceptible to low-frequency users and spammers. Low-frequency
users post only few documents on a social network. In the extreme case, a user only posts a single document about a
topic and thus receives a high qualification score of $1$ and a perfect interest score of $1$. Thus, we cannot safely
select these users as experts. Spammers on the other hand are
users that post many documents about the same topic to pretend to have high qualification and interest. 

Pal and Counts' measure against low-frequent users and spammers is to use a probabilistic clustering approach (Gaussian
mixture model) \cite{pal2011identifying}. In our case, because we want to combine expertise features with social
features later, we will use a different approach. We exclude low-frequency users by smoothing the qualification and
interest features. Smoothing assigns more weight to users that posted more documents on a topic, but not necessarily
completely about the topic. The modified formulas of smoothed qualification and interest are as follows, where $\mu$ is
the average mean of the non-smoothed ratio and interest respectively and $N$ is the number of data points:
\begin{align*}
K_1'(c,u,t) &= \frac{\text{CA}_{(c,u,t)}+ \mu}{\text{A}_{(c,u,t)}+ N} + \frac{\text{OP}_{(c,u,t)} + \mu}{\text{P}_{(c,u,t)} + N} \\
K_2'(c,u,t) &= \frac{\text{P}_{(c,u,t)} + \mu}{\text{P}_{(c,u)} + N} 
\end{align*}
Smoothing in our approach is similar to additive smoothing or Laplace smoothing \cite{manning2008introduction}. It
assigns more weight to users that post more documents on topic compared to low-frequency users that post only few
documents on topic. Thus, we effectively exclude low-frequency users from being selected as experts.

Even with smoothing, spammers will receive high qualification and interest scores. We exclude spammers from being
selected as experts by disregarding users with low values in at least one of the expertise dimensions (qualification and
interest) or social dimensions (responsiveness and activity, explained later). Our hypothesis is that although spammers
have high qualification and interest scores, they have low responsiveness and/or activity scores (i.e., do not
communicate with other users in a social network). Judging from the experience with our social routing algorithm, this
hypothesis holds true for the social networks we analyzed. Thus, we effectively exclude spammers from being asked
questions by our social routing algorithm.

The next subsection explains how to combine expertise detection with social features. The goal is to find experts who
are likely to answer questions about their topics of expertise. For this, we will introduce features that describe the
experts' communication behavior on social networks.

\subsection{Social availability}

While it is clear that \emph{Knowledge} has to be closely related to a topic, we noticed that also \emph{Availability}
works in the same way. For the same level of expertise, people show different response rates on different arguments due
to social trends, personal preferences, or temporal convenience. Furthermore, by computing \emph{Activity} on distinct
subjects it is possible to give users the option to contribute on a diverse range of questions instead of overloading
them on the same topic. Formally, we compute
responsiveness ($A_1$) and activity ($A_2$) for a user $u$ in a crowd $c$ on a topic $t$
through the following definitions. Table~\ref{tab:socialmetrics} explains the terms used in the definition.
\begin{align*}
A_1(c,u,t) &= \frac{\text{AQ}_{(c,u,t)} + \mu}{\text{PQ}_{(c,u,t)} + N} + \frac{\text{CP}_{(c,u,t)} + \mu}{\text{P}_{(c,u,t)} + N} +  \frac{1}{\text{RT}_{(c,u,t)}}\\
A_2(c,u,t) &= now - \max\Big(time(LQ_{(c,u,t)}),time(LA_{(c,u,t)}) \Big)
\end{align*}
\begin{table}[t]
\caption{Metric definitions of social availability.}
\label{tab:socialmetrics}
\centering
\begin{tabular}{@{}p{2cm} l@{}}
  \toprule
  Metric & Definition \\ \midrule
  CP     & conversational post \\
  PQ     & question presented to the user \\
  AQ     & question answered by the user \\
  RT     & average response time \\
  LQ     & last question presented \\
  LA     & last answer provided \\
  \bottomrule
\end{tabular}
\end{table}
Note that \emph{Responsiveness} captures the responsiveness of the user to our tasks as well as to posts initiated by
other users in the network. At the same time, it also includes the average response time on the topic. \emph{Activity}
then keeps track of the last Q\&A event with the user on the topic. This means that a user that was recently asked on
a topic will not be accessed on the same topic any time soon, yet he might still be a good candidate for other topics
on which he is currently idle. In general, it is useful to take into account the current status of the user in the
network (online or offline) because there is a higher chance that users who are currently online will respond. We did
not investigate this because of privacy concerns that could make the users reluctant to contribute during the bootstrapping phase of this project. 

\begin{table}[t]
\caption{Refresh rate and window size for feature collection.} 
\centering
\label{tab:feature_analysis}
\begin{tabular}{@{}lll@{}}
  \toprule
  Feature         & Refresh rate  & Window size \\ \midrule
  Qualification   & Daily         & Preferably forever \\
  Interest        & Daily         & Monthly \\
  Responsiveness  & Immediately   & Monthly \\
  Activity        & Immediately   & Weekly \\ \bottomrule
\end{tabular}
\end{table}

The routing strategy described in the next section requires that the underlying features are up to date so that it can
correct itself and not select users which do not collaborate in Q\&A. From our observations it results that the social
\emph{Availability} features tend to change much faster than the \emph{Knowledge} ones. Thus, we update them every time
a question is asked through CrowdSTAR or answered by the users, while \emph{Knowledge} is updated daily. The window size
of the data that needs to be considered for each feature is also an important factor since some of the features have a
higher longevity than others. For example, \emph{Qualification} is a more long-term feature than \emph{Interest}
assuming that certain skills do not deteriorate drastically over time and that interests are more likely to change.
Table~\ref{tab:feature_analysis} shows a summary analysis of these two factors (refresh rate and window size) for each
of the features in the proposed user model. All features were separately computed for both networks over a one-month time window and
were then input into the social task router described in the next section.

%% file: sections/sec5_socialrouting.tex
\section{Social Task Routing}
\label{sec:taskrouting}
\begin{figure}[t]
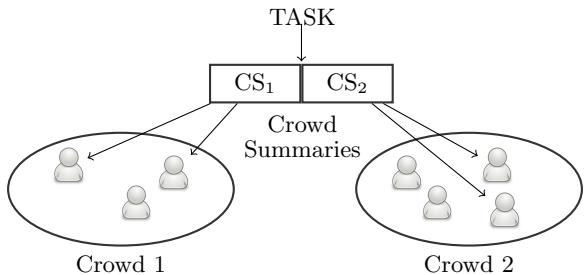
 
	\centering
	\begin{tikzpicture}
		\tikzstyle{texts}=[draw=none,fill=none]
		\tikzstyle{ell}=[ellipse, minimum width = 30mm, minimum height = 15mm, thick, draw =black!80, node distance = 16mm]
		\tikzstyle{rec}=[rectangle, minimum width = 12mm, minimum height = 5mm, thick, draw =black!80, node distance = 16mm]
		\tikzstyle{line}=[draw]
		\node[ell] (C1) [, label=below:{Crowd 1}] { };
		\node[inner sep=0pt] (U1) at (0.2, -0.2){\includegraphics[width=0.5cm]{user}};
		\node[inner sep=0pt] (U2) at (0.7, 0.2){\includegraphics[width=0.5cm]{user}};
		\node[inner sep=0pt] (U3) at (-0.7, 0.3){\includegraphics[width=0.5cm]{user}};
		\node[ell] (C2) [right=1.6 cm of C1, label=below:{Crowd 2}] { };
		\node[inner sep=0pt] (U4) at (4.2, -0.2){\includegraphics[width=0.5cm]{user}};
		\node[inner sep=0pt] (U5) at (3.8, 0.2){\includegraphics[width=0.5cm]{user}};
		\node[inner sep=0pt] (U6) at (5.1, -0.3){\includegraphics[width=0.5cm]{user}};
		\node[inner sep=0pt] (U7) at (5.0, 0.3){\includegraphics[width=0.5cm]{user}};
		\node[rec] (CS1) [above right=0.6 cm and 0.1 cm of C1, label=below:{}] { $\text{CS}_1$};
		\node[rec] (CS2) [right=0.0 cm of CS1, label=below:{}] {$\text{CS}_2$};
		\path [line] (CS1) [->] -- (U2);
		\path [line] (CS1) [->] -- (U3);
		
		\path [line] (CS2) [->] -- (U7);
		\path [line] (CS2) [->] -- (U6);
		\node[texts] (S1) [below right =0.5 cm and -0.1 cm of CS1, label=above:{Crowd}] { }; \node[texts] (S2) [below =0.1 cm of S1, label=above:{Summaries}] { }; \node[texts] (T) [above =1.2 cm of S1, label=above:{TASK}] { }; \draw (2.4,2.2) [->]
-- (2.4,1.7);
	\end{tikzpicture}
	\caption{Task routing.}
	\label{fig:Routing}
\end{figure}

According to Law and von Ahn~\cite{von2009human} there exist two forms of assigning tasks to crowd members, referred to
as push and pull approaches. Pull approaches let the users select the tasks, while the push approaches explicitly match
the tasks to users. Many paid and non-paid platforms of crowdsourcing choose to adopt a hybrid model of these two forms
so that the self-initiative of the contributors can be leveraged to increase the quality of the results. In our work,
the social task routing belongs to the second form of task assignment but CrowdSTAR design is aware of the
self-regulating events that happen in dynamic crowds where members can make free choices. The responsibility of the
social task router in our system is to perform two kinds of routing in the following order: (1) routing across the
crowds and (2) routing within a crowd. The first one forwards the task to one of the candidate crowds, while the second
continues the routing by pushing the question to individual members from the crowd chosen in Step~1.
Figure~\ref{fig:Routing} depicts an overall picture of the process.

\subsection{Routing tasks within a crowd}

In order to consider all the features in the user utility model, the routing algorithm needs to explore the possible
trade-offs between the features and access only those users which appear to dominate the rest of the crowd. For this
purpose we select as a candidate user set the group of users which is not dominated by others in at least one of the
dimensions. We refer to this candidate set as the \emph{crowd skyline} for the topic associated to the task. Figure
\ref{fig:Skyline} illustrates a sample output for two dimensions where the connected points represent the crowd skyline.
Depending on the topic, the crowd expertise and how much redundancy one wants from the crowd, it can happen that the
number of users in the skyline is not enough. In Quora for example, it is quite common that there exist a very few
active users that dominate all the others which would prevent us from involving other users. For this purpose, we decide
to continue running the skyline algorithm even beyond the first skyline. For example, in Figure~\ref{fig:Skyline} the
data points connected by the dashed line represent the second skyline.

The skyline computation uses the algorithm introduced by Kossmann et al.~\cite{kossmann2002shooting}. It applies a
recursive nearest-neighbor search that continuously prunes from the search space regions that are dominated by the
actual best data point not yet included in the skyline. The algorithm has a good pruning rate which is a necessary
property for our routing algorithm to scale. Figure~\ref{fig:pruning} shows the pruning rate for the set of the 300,000
most active users on Twitter on the popular topic \emph{music}. The size of the set of the remaining candidate users decreases
drastically with every iteration, making each iteration faster to compute. Furthermore, a good property of the algorithm
is the early output of skyline points, which is useful for very large data when it is not possible to wait until the
whole computation finishes. It is possible and reasonable to start routing questions to users while the skyline
computation is still running. We further prune the search space by disregarding users which have very low values in at
least one of the axes (the dashed regions in Figure~\ref{fig:Skyline}) because our experiments showed that these regions
contain mostly spammy and non-responsive accounts.
\begin{figure}[t]
	\centering
	\begin{tikzpicture}
		\tikzstyle{texts}=[draw=none,fill=none]
		\tikzstyle{ell}=[ellipse, minimum width = 30mm, minimum height = 15mm, thick, draw =black!80, node distance = 16mm]
		\tikzstyle{rec1}=[rectangle, minimum width = 34mm, minimum height = 4mm, thin, draw =black!50, node distance = 16mm, pattern=north east lines]
		\tikzstyle{rec2}=[rectangle, minimum width = 4mm, minimum height = 29mm, thin, draw =black!50, node distance = 16mm, pattern=north east lines]
		\tikzstyle{line}=[draw]
		\draw[thick] (0,0) [->] -- (3.5,0);
		\draw[thick] (0,0) [->] -- (0,3);
		\node[texts] (A) at (2.0, -0.76) [label=above:{Availability}] { };
		\node[texts] (K) at (-0.25, 0.5) [label=above:{\begin{turn}{+90}Knowledge\end{turn}}] { };
		\foreach \Point in {(2,2), (1,1), (2,2.5), (1,1.5), (0.1,2.7), (0.1,2.7), (2.1,1.4),  (2.7,1.9) , (3.1,1.4), (3.2,0.2), (2.1,1), (1.5,2.2), (0.2,0.6), (0.8,2.7), (2.8, 1.3), (2.9, 1.1)}{
    			\node at \Point {$\circ$};
		}
		\foreach \Point in {(0.8,2.7), (2,2.5), (2.7,1.9), (3.1,1.4)}{
    			\node at \Point {\textbullet};
		}
		\draw [densely dotted] (2.8, 1.3) -- (2.9, 1.1);
		\draw [densely dotted] (2,2) -- (2.8, 1.3);
		\draw [densely dotted] (2,2) -- (1.5,2.2);
		\draw (0.8,2.7) -- (2,2.5); 
		\draw (2.7,1.9) -- (2,2.5); 
		\draw (2.7,1.9) -- (3.1,1.4); 
		\node[rec1] (R1) [] at (1.7,0.2){}; 
		\node[rec2] (R2) [] at (0.2, 1.45){};
	\end{tikzpicture}
	\caption{Example of the crowd skyline in two dimensions.}
	\label{fig:Skyline}
\end{figure}
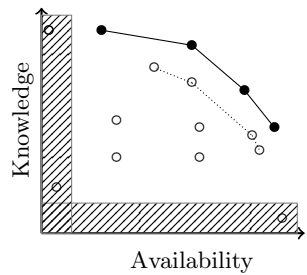
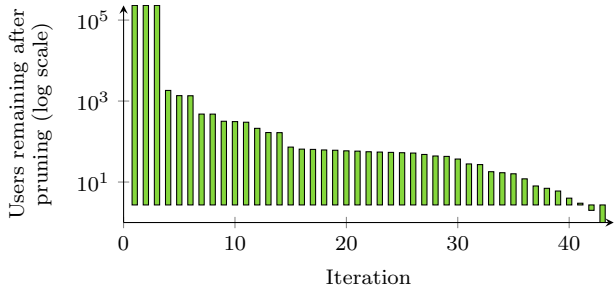
\begin{figure}[t]
  \pgfplotstableread{graphdata/skyline_prune.dat}{\prune}
  \definecolor{gg2}{RGB}{133, 215, 60}
  \begin{tikzpicture}
  \small
  \begin{axis}[
    xlabel style={align=center},
    xlabel=Iteration,
    ylabel style={align=center},
    ylabel=Users remaining after \\ pruning (log scale),
    width=0.96\columnwidth,
    height=.53\columnwidth,
    ymode = log,
    bar width=2pt,
    axis lines=left,
    xmin = 0,
    ymin = 0,
    xmax=44]
  \addplot[ybar,fill=gg2] table [x={it}, y={no}] {\prune};
  \end{axis}%
  \end{tikzpicture}
  \caption{Pruning rate of the skyline computation.}
  \label{fig:pruning}
\end{figure}

Whenever the user utility model is updated, the crowd skyline needs to be recomputed. Different users may appear in the
skyline. For instance, if a user has just answered a question, the respective \emph{Activity} is going to be updated with
a very low value excluding this way the user from the candidate set to ask. Similarly, if someone gradually changes
\emph{Interest} from photography to video and starts posting and answering more on the latter topic, the same switch
will happen to his or her membership in the topic skyline.

In our routing experiments we did not ask all the users in the set since this would be too intrusive. Instead, we start
in the middle of the skyline and then incrementally move towards the edges of the skyline in both directions. However,
exploring different segments of the skyline is also effective because it gives a chance of participation and improvement
to users that do not have the highest scores in all dimensions. Indeed, as we show in the experiments section, such
users exist in both crowds that we studied.

\subsection{Routing tasks across multiple crowds}

The decision of crowd selection is based on an aggregate summary of each crowd. Although we index the features of all
users, we do not use all of them to build the crowd summary shown in Figure \ref{fig:Routing}. The summary includes only
those members which will possibly be considered for question asking in the near future, i.e. the crowd skyline. The
following formulation defines the summary of a crowd $c$ on an arbitrary feature $f$ for a topic $t$.
\begin{align*}
\text{Summary}(c,t,f) &= \frac{\sum_{u \in \text{skyline}(c,t)}{f(c, u, t)}}{|\text{skyline}(c,t)|}
\end{align*}
Having the summary on each dimension, the final crowd score of the crowd on the topic can be computed as a weighted
linear combination of all the features. Note that \emph{Activity} is excluded from the final score given that it is
closely related to individual users and should not affect the overall accessibility of the crowd.
\begin{align*}
\text{Score}(c,t) &= \sum_{f \in \{K_1, K_2, A_1\}} \big(  w_f \cdot \text{Summary}(c,t,f)\big)
\end{align*}
Assigning different weights to the dimensions allows for adapting the routing algorithm to the task requirements. For
example, if one is interested in solving a survey task, the highest weights should go to \emph{Interest} and
\emph{Responsiveness} considering that the crowd members will only give their personal opinion and not actually solve a
problem. For a fair comparison between crowds the number of users in the skyline of each crowd should be balanced which
is very unlikely to happen given the different feature distributions. This problem is solved by choosing for both crowds
an equal number of points as skyline representatives and possibly making use of the lower-level skylines as defined
previously.

In practice, we observed that Twitter and Quora have similar scores for very popular general topics like \emph{music},
\emph{sport} or \emph{travel}. On the other hand, they have significantly different scores for topics that mostly fall
in the domain of only one of the crowds. Such examples are \emph{startup}, \emph{silicon valley} for Quora, and
\emph{nfl}, \emph{golden globe} for Twitter. For this reason, we apply a more relaxed access distribution mechanism
across crowds. More specifically, given a specific budget $B$ (the number of people we want to ask the same question for
redundancy purposes), we decide to equally distribute the budget if the relative difference between the summary scores
is less than 25\%. Otherwise, the budget is distributed proportionally.

%% file: sections/sec6_crowdstar.tex
\section{CrowdSTAR System}
\label{sec:crowdstar}
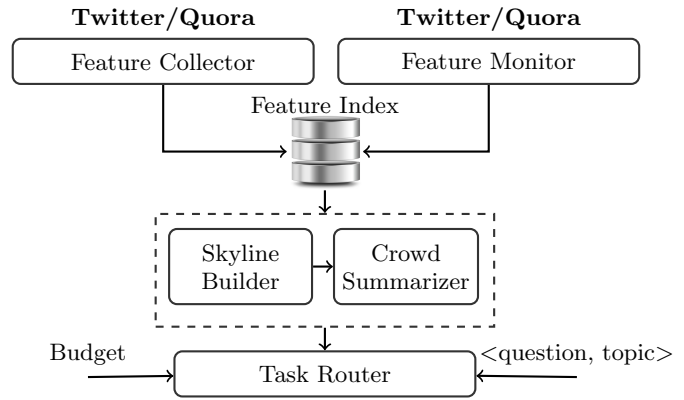
\begin{figure}[t]
	\centering
	\begin{tikzpicture}[every text node part/.style={align=center}]
		\tikzstyle{texts}=[draw=none,fill=none]
		\tikzstyle{ell}=[ellipse, minimum width = 30mm, minimum height = 15mm, thick, draw =black!80, node distance = 16mm]
		\tikzstyle{rec}=[rectangle, minimum width = 40mm, minimum height = 6mm, thick, draw =black!80, node distance = 0mm, rounded corners=3pt]
		\tikzstyle{line}=[draw]
		\node[rec] (T) {Feature Collector};
		\node[texts] (TL) [above=-0.25 cm and 0 cm of T, label=above:{\textbf{Twitter/Quora}}] { };
		\node[rec] (Q) [right=0.3 cm of T] {Feature Monitor};
	    \node[texts] (QL) [above=-0.25 cm and 0 cm of Q, label=above:{\textbf{Twitter/Quora}}] { };
		\node[inner sep=0pt] (I) at (2.15, -1.2){\includegraphics[width=1.0cm]{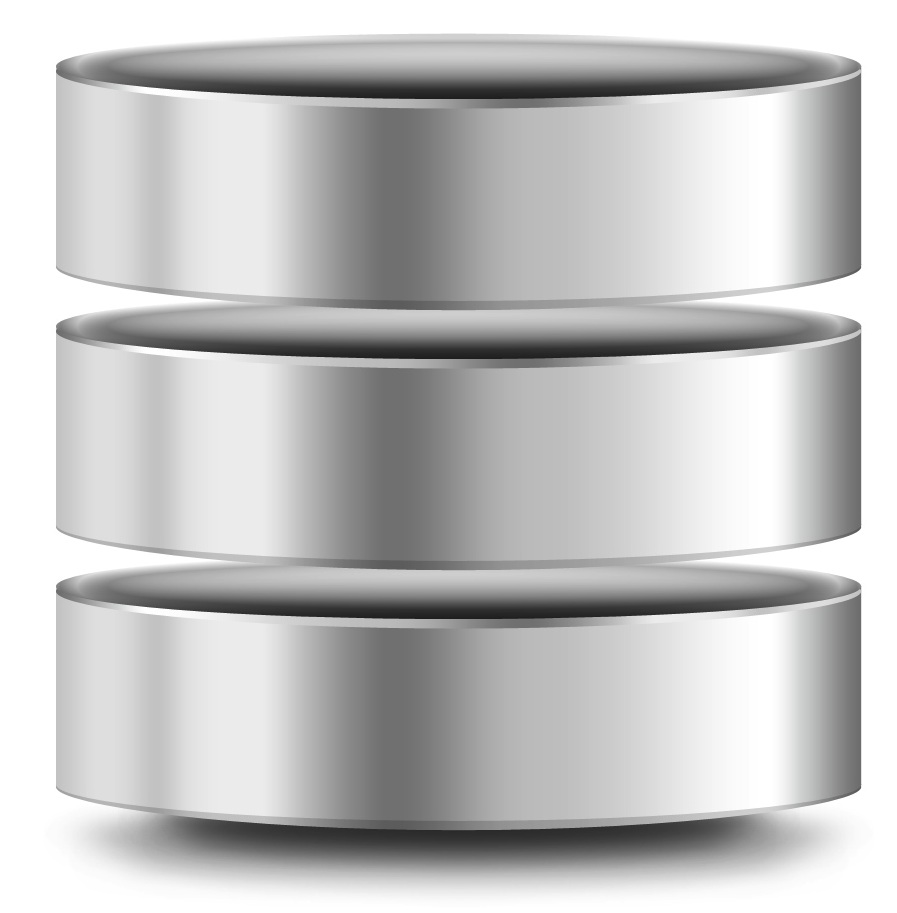}};
		\node[texts] (IL) [above=-0.35 cm and 0 cm of I, label=above:{Feature Index}] { };
		\draw[thick][->] (T) |- (I);
		\draw[thick][->] (Q) |- (I);
		
		\node[rectangle, dashed, minimum width = 45mm, minimum height = 15mm, thick, draw =black!80, node distance = 0mm] (S) [below=0.3 cm of I] {};
		\node[rectangle, minimum width = 19mm, minimum height = 10mm, thick, draw =black!80, node distance = 0mm, rounded corners=3pt] (S1) [below left=0.5 cm and -0.35 cm of I] {Skyline \\Builder};
		\node[rectangle, minimum width = 19mm, minimum height = 10mm, thick, draw =black!80, node distance = 0mm, rounded corners=3pt] (S2) [below right=0.5 cm and -0.4 cm of I] {Crowd \\Summarizer};
		\draw[thick][->] (S1) -- (S2);
		\draw[thick][->] (I) -- (S);
		\node[rec] (P) [below=0.3 cm of S] {Task Router};
		\draw[thick][->] (S) -- (P);
	    \node[texts] (B) at (-1, -3.9) {Budget};
	    \draw[thick][->] (-1, -4.2) -- (P);
	    \node[texts] (QT) at (5.5, -3.9) {$<$question, topic$>$};
	    \draw[thick][->] (5.5, -4.2) -- (P);
	\end{tikzpicture}
	\caption{CrowdSTAR architecture.}
	\label{fig:CrowdStarSystem}
\end{figure}

In the CrowdSTAR system we put together all the functionalities that have been described so far.
Figure~\ref{fig:CrowdStarSystem} depicts the main modules of the system and how they interconnect with each other. All
the modules are implemented in C\# and ran on SCOPE \cite{chaiken2008scope} on a large computing cluster.

\subsection{Components}
\noindent \emph{\textbf{Feature Collector}} gathers the textual evidence of users' expertise on the topics that we
experimented with and for the two social networks we are considering, Twitter and Quora. The task of the module is to
fetch those metrics from Table~\ref{tab:expertise_metrics} and~\ref{tab:socialmetrics} that are part of the user's
activity in the social network regardless of their participation in our tasks. Such activities in Quora for example are
native answers, questions, comments, and blog posts. This part of the metrics was very important during the
bootstrapping stage of the project and appeared to be useful in the other stages as well for identifying new candidate
experts. The module collects the data of the latest one month interval and forwards them to the \emph{Feature Index}
module.

\noindent \emph{\textbf{Feature Monitor}} As described in the previous sections, the user utility model also makes use
of the activity of the users in our system. This module monitors in real time whether the posted questions have been
answered from the targeted users and sends this information to the \emph{Feature Index}. The respective instances of
this module are continuously running since some of the data that they collect are time-critical, particularly because of
their contribution to availability and activity. Technically, we use the Twitter API and Quora RSS Feeds for our
representative accounts in each of the networks.

\noindent \emph{\textbf{Feature Index}} Having the incoming data from the previous modules, the \emph{Feature Index}
recomputes the changed dimensions of the user utility model. Since all of the metrics consist of parallelizable
aggregation functions, the computation can be easily carried out in any architecture that supports MapReduce tasks
\cite{chaiken2008scope}. At the moment, the index keeps track of approximately the top 300,000 active users in Twitter
and top 45,000 users in Quora.

\noindent \emph{\textbf{Skyline Builder}} and \emph{\textbf{Crowd Summarizer}} As soon as the \emph{Feature Index} is
updated on a certain topic, the \emph{Skyline Builder} gets updated on the same topic by recomputing the skyline which
is then used to refresh the crowd summary scores in the \emph{Crowd Summarizer}. Although in our experiments the skyline
generation is not a bottleneck, for larger scale applications it is possible to adapt the algorithm proposed in
\cite{kossmann2002shooting} for MapReduce frameworks by first computing in parallel the skyline candidates for vertical
cuts of the feature spaces and then merge them into a single one. A survey and other solutions on parallel skyline
queries is presented in \cite{wu2006parallelizing}.

\noindent \emph{\textbf{Task Router}} Given a question task associated to a main topic, the \emph{Task Router}
incrementally routes the question according to the summarized crowd scores and the topical skylines. The posting process
is done through the Twitter API while Quora does not provide an API yet and needs manual question posting. The budget
here refers to the number of users to be asked as a degree of crowdsourcing redundancy that can be specified by the end
user. The \emph{Task Router} can be further extended in the future to better support individual characteristics of each
crowd. For example, in Quora each member has a self-assigned price in virtual credits for each question targeted to him
or her. Thus, the asking ``cost'' and the budget can also be measured in credits.

\subsection{Use Cases and Applications}

The proposed functionalities of CrowdSTAR are designed to help end users trying to solve challenging tasks with the help
of human power available on the Web. We identify two main use cases for such a system:
\begin{enumerate}
\item \textbf{Peer-to-peer routing (UC1)}. In this use case the system is used to propose to the asker a set of candidate experts in various networks.
 Afterwards, the asker can freely choose how many and which of the presented users to ask and then can directly contact them through CrowdSTAR.
 In this scenario the identity of both peers is made public to each other. The model is used by most of the Q\&A platforms available today and works
 well for single communities.
\item \textbf{Answer provider (UC2).} This option implies that both of the parties (the asker and the responder) remain anonymous. The offered service 
not only finds the possible experts but also contacts them on behalf of the askers and then sends back the answers. Being more discrete, this model 
does not imply many of the privacy issues that come with the first use case and also reduces the latency of the end-to-end process.
\end{enumerate}

%% file: sections/sec7_experiments.tex
\section{Experimental Evaluation}
\label{sec:experiments}
In this section we discuss how we collected the data for populating the user utility model and how we performed the
routing experiments across and within Twitter and Quora.

\subsection{Feature collection}

The features of each user were computed from the most recent one month interval. We focused on a broad range of topics
(35 in total) from domains like technology, hobbies, news, and entertainment. In Quora we consider that a post falls
within a topic if this is claimed from the author or Quora's maintenance staff since the mapping is highly accurate in
this network. The posts in Twitter are not as structured. Thus, we categorize a tweet within a topic if the topic word
explicitly appears in the tweet text. Involving the topic-to-topic relationships would result into misleading outcomes.
For example, a user who talks about \emph{soccer} may not be an expert in \emph{sport} and vice versa.

Table~\ref{tab:features} shows an example of retrieving the top five users in Twitter with respect to
\emph{Qualification} and \emph{Responsiveness} for topic \emph{hiking}. Note that the most qualified users are famous
accounts on the topic but not necessarily personal accounts, while the most responsive ones match to people who tend to
answer and converse more on \emph{hiking}. They are still knowledgeable but their attention is not focused on a single
interest only. A similar phenomenon can be observed in Quora but with slightly different motivations. The accounts in
this network are not commonly used for company advertisements or spam but rather for self promotion. Here we identify
two different categories of accounts: those who tend to create a large number of relatively short and mid-quality posts
and those who prefer posting less content but of a higher quality. The quality of a post or an answer can be verified
through the number of upvotes that other users assigned. The upvotes also can be interpreted as the visibility or the
impact of the content in the network. Due to these reasons, we include them in the qualification computation by weighing
each post and answer with its upvote. After this adaptation, we are able to distinguish between the two categories of
users mentioned before. The same adaptation is technically possible in Twitter as well by analogously making use of the
number of times the tweet is marked as favorite. Unfortunately, this information is most of the times not available in
Twitter since the visibility of the user-to-user tweets is very low.

\begin{table}[t]
		\caption{Example of top 5 Twitter users wrt. to qualification and responsiveness for topic \emph{hiking}.}
	\centering 
	\label{tab:features} 
	\begin{tabular}{@{}ll@{}} 
		\toprule 
		\multicolumn{2}{c}{TOP 5 WHERE topic='hiking' ORDER BY}\\
		\midrule
		QUALIFICATION & RESPONSIVENESS\\
		\midrule
		@hiking\_camping& @thatoutdoorguy\\
		@letsgoforahike& @nickandriani\\
		@mightycrack & @astrogerly\\
		@etravelhotels & @melissabravery\\
		@outdoorgeardotd & @rsrigda\\
		\bottomrule		
	\end{tabular}
\end{table}

\begin{figure}[th]
\pgfplotstableread{graphdata/twitter_travel.dat}{\travel}
\pgfplotstableread{graphdata/quora_travel.dat}{\quoratravel}
\pgfplotstableread{graphdata/twitter_hiking.dat}{\hiking}
\pgfplotstableread{graphdata/quora_hiking.dat}{\quorahiking}
\centering \begin{tikzpicture}[font=\small] \small
	\begin{axis}
	[xlabel style={align=center}, xlabel=,
	ylabel style={align=center}, ylabel=\textbf{Twitter} \\ Responsiveness,
    width=.53\columnwidth,
    height=.53\columnwidth,
    ytick={0, 0.2,0.4,0.6,0.8},
	xticklabel style={/pgf/number format/.cd,fixed,precision=2}
    ]
	\addplot[scatter, only marks, scatter src=x+y] table [x={qualification}, y={responsiveness}, z={sum}] {\travel};
	\end{axis}
	\node[rectangle, densely dotted, minimum width = 8mm, minimum height = 12mm, thin, draw =black!80] (R1) at (2.2, 2.2) {};
\end{tikzpicture}
\begin{tikzpicture}[font=\small]
	\small
	\begin{axis}
	[xlabel style={align=center}, xlabel=,
	ylabel=,
    width=.53\columnwidth,
    height=.53\columnwidth,
    ytick={0, 0.2,0.4,0.6,0.8},
	xticklabel style={/pgf/number format/.cd,fixed,precision=2}
    ]
	\addplot [scatter, only marks, scatter src=y+y+y+x] table [x={qualification}, y={responsiveness}] {\hiking};
	\end{axis}
\end{tikzpicture}
\begin{tikzpicture}[font=\small]
	\small
	\begin{axis}
	[xlabel style={align=center}, xlabel=Qualification \\ \emph{travel},
	ylabel style={align=center}, ylabel=\textbf{Quora} \\ Responsiveness,
    width=.53\columnwidth,
    height=.53\columnwidth,
    ytick={0, 0.2,0.4,0.6,0.8},
	xticklabel style={/pgf/number format/.cd,fixed,precision=2}
    ]
	\addplot [scatter, only marks, scatter src=x+y] table [x={qualification}, y={responsiveness}] {\quoratravel};
	\end{axis}
	\node[rectangle, densely dotted, minimum width = 12mm, minimum height = 6.5mm, thin, draw =black!80] (R2) at (2.15, 2.47) {};
\end{tikzpicture}\hspace*{6pt}
\begin{tikzpicture}[font=\small]
	\small
	\begin{axis}
	[xlabel style={align=center}, xlabel=Qualification \\ \emph{hiking},
	ylabel=,
    width=.53\columnwidth,
    height=.53\columnwidth,
    ytick={0, 0.2,0.4,0.6,0.8},
	xticklabel style={/pgf/number format/.cd,fixed,precision=2}
    ]
	\addplot [scatter, only marks, scatter src=x+y] table [x={qualification}, y={responsiveness}] {\quorahiking};
	\end{axis}
	\node[rectangle, densely dotted, minimum width = 3.8mm, minimum height = 3.8mm, thin, draw =black!80] (R1) at (2.02, 2.63) {};
\end{tikzpicture}
\caption{Qualification and Interest scores for questions on \emph{travel} and \emph{hiking} on Quora and Twitter.}
\label{fig:QuoraTwitterTopics}
\end{figure}
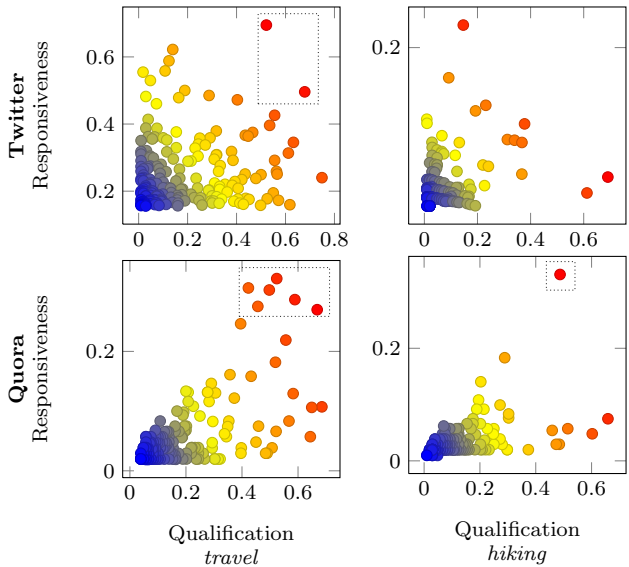

In Figure~\ref{fig:QuoraTwitterTopics} we show the \emph{Qualification} and \emph{Responsiveness} for 200 most active
users of both networks for the topics \emph{travel} and \emph{hiking}. Users of the same color gradient would belong to
the same skyline level as defined in our method. As
expected, there is not necessarily a strong
correlation between them (also the case for the other features) which supports once again the fact that using a linear
combination or a generalization of all the features (e.g., the total number of posts) is less informative and that the
identified dimensions in the user utility model are present in real-world data. User data points of this nature, but of
a larger scale, serve as an input for the social task routing algorithm. Ideally, we would like to choose only
points that have very good scores on all the features such as those that fall within the dotted rectangles in the
figure. In practice, this is not always feasible. For example, comparing
the graphs for the two topics we can understand that the skyline region is more dense for popular and general topics
like \emph{travel}. For more specific ones like \emph{hiking}, especially on Quora we can notice the existence of very
few dominating experts that are in fact very well-known from other members with common interests in the same community.
Also the user points in Twitter are more scattered as this network is larger, more diverse and less structured compared
to Quora.

\subsection{Question posting}
\begin{table*}[t]
\caption{Comparison of Twitter and Quora task routing support.}
\centering
\label{tab:crowd_comparison}
\begin{tabular}{@{}p{3.5cm} @{}p{5cm} @{}p{6cm}}
  \toprule
   		 					& \textbf{Quora} 					& \textbf{Twitter}\\ \midrule
  Responsiveness      		& 64\% 								& 44\% 		\\ \hline
  Questions answered      	& 85\% 								& 44\% 		\\ \hline
  Average response time     & $\sim$24 hours (1st response)		& 12.7 hours\\ \hline
  Accuracy      			&  \multicolumn{2}{c}{80\%-90\% (manually evaluated)} \\ \hline
  Asking tone     			& Formal 							& Informal \\ \hline
  Question visibility     	& Many users						& Mainly the assigned user\\ \hline
  \#Users answering     	& Many users						& Only the assigned user\\ \hline
  Human intervention     	& Thanking (built-in)				& Introduction\\
  					     	& 									& Greeting\\
							& 								    & Thanking\\\hline
  Question length     		& Not restricted & 140 char max\\\hline
  Answer properties     	& Long	 							& 140 char max \\ 
  					     	& Elaborated					 	& Concise\\ \hline
  Quality control    		& Upvotes and Editors				& None available\\ 
    			    		& 									& Possible candidates: \#retweets, \#favorites\\ \hline
  Types of Q\&A      		& Recommendation multiple items		& Recommendation single item\\ 
        					& "How to'' explanations 			& Laconic explanations \\ 
      						& Only interesting surveys			& Survey\\ 
      		      			& 									& Factual \\
  \bottomrule
\end{tabular}
\end{table*}

\begin{figure}[th]
	\centering
	\begin{tikzpicture}
		\tikzstyle{texts}=[draw=none,fill=none]
		\tikzstyle{ell}=[ellipse, minimum width = 30mm, minimum height = 15mm, thick, draw =black!80, node distance = 16mm]
		\tikzstyle{rec}=[rectangle, minimum width = 30mm, minimum height = 5mm, thick, draw =black!80, node distance = 0mm, rounded corners=3pt]
		\tikzstyle{line}=[draw]
		\node[rec] (T) {question \#ask \#topic};
		\node[rec] (S1) [below left=0.3 cm and -0.5 cm of T] {Introduce project};
		\node[rec] (S2) [below right=0.3 cm and -0.5 cm of T] {"Hi!"};
		\node[rec] (Q1) [below=0.35 cm of S1] {Post question};
		\node[rec] (Q2) [below=0.35 cm of S2] {Post question};
		\node[rec] (TH) [below=2.0 cm of T] {"Thank you!"};
		\path [line] (T) [->] -- (S1);
		\path [line] (T) [->] -- (S2);
		\path [line] (S1) [->] -- (Q1);
		\path [line] (S2) [->] -- (Q2);
		\path [line] (Q1) [->] -- (TH);
		\path [line] (Q2) [->] -- (TH);
	\end{tikzpicture}
	\caption{Alternative asking strategies in Twitter.}
	\label{fig:TwitterAsking}
\end{figure}
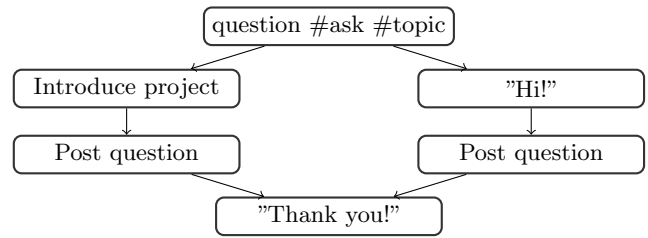

\begin{figure}[th] \centering
	\begin{tikzpicture}
		\tikzstyle{texts}=[draw=none,fill=none]
		\tikzstyle{recQ}=[rectangle, minimum width = 70mm, minimum height = 5mm, thick, draw =gray!50, node distance = 0mm, rounded corners=1pt]
		\tikzstyle{recT}=[rectangle, minimum width = 70mm, minimum height = 5mm, thick, draw =gray!50, node distance = 0mm, rounded corners=1pt]
		\node[inner sep=0pt] (Q) at (0, 0){\includegraphics[width=0.5cm]{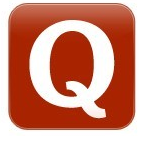}};
		\node[below=4.5 mm of Q, inner sep=0pt] (T) at (0, 0){\includegraphics[width=0.5cm]{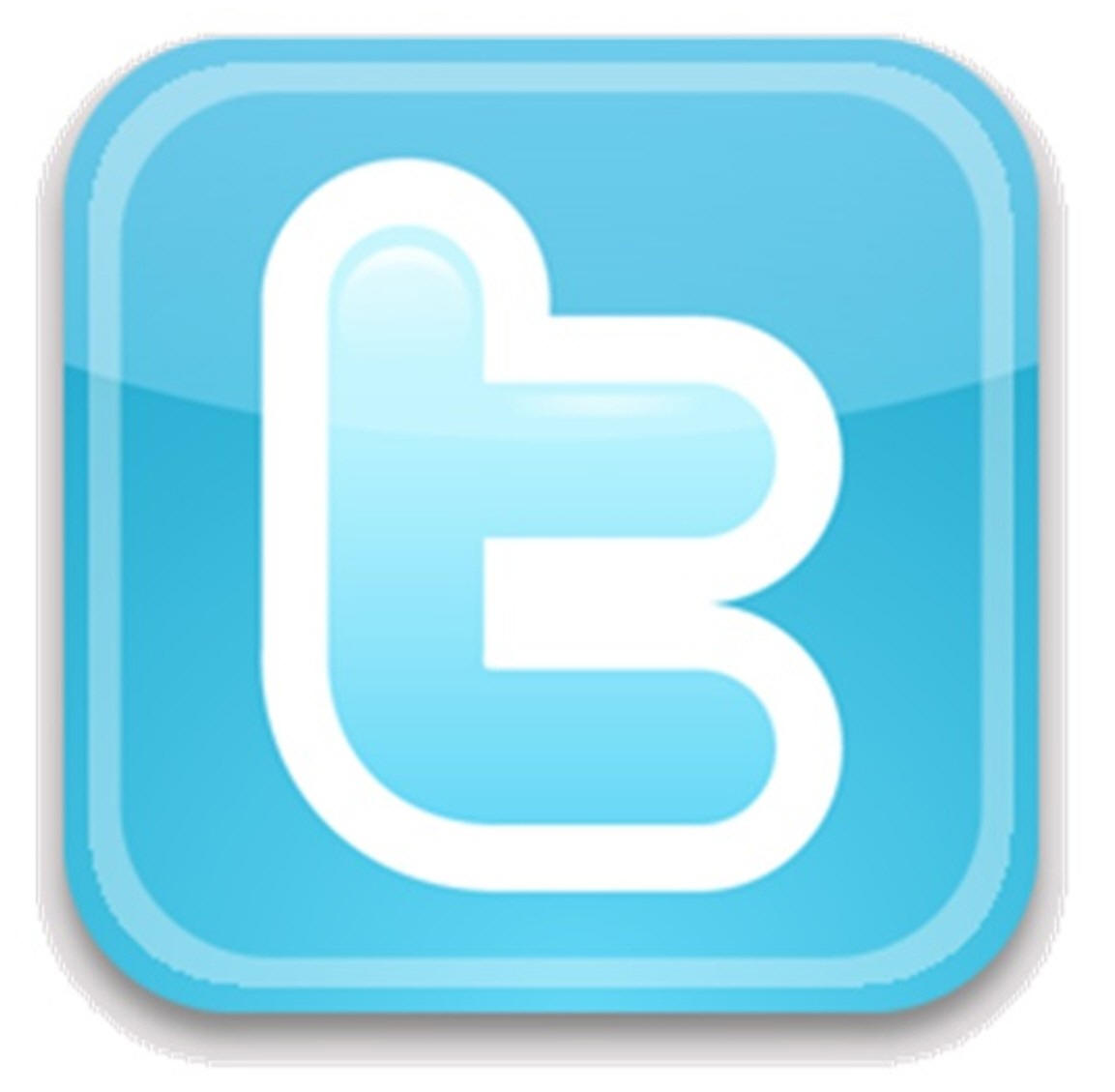}};
		\node[recQ] (QQ) [right=1.5 mm of Q, text width=77mm] {How popular is motorbiking among women in USA?};
		\node[recT] (QT) [below=1.5 mm of QQ, text width=77mm] {Hi \emph{@user}! Do you think motorbiking is popular \\ among women in USA? \#ask \#motorbiking};
	\end{tikzpicture}
	\caption{Example of asking the same question in Twitter and Quora.}
	\label{fig:Questions}
\end{figure}

For the purpose of conducting question routing experiments we created two different accounts in Twitter and Quora named
respectively \emph{@SocialQARouting} (\url{https://twitter.com/SocialQARouting}) and \emph{Ada Floyd} (\url{http://www.quora.com/Ada-Floyd}). Both of them were first bootstrapped by gradually asking
questions and posting other non-asking content. In Twitter we alternated two asking strategies as shown in
Figure~\ref{fig:TwitterAsking} and also attached the \#ask and \#$<$topic$>$ hashtags to the question text to increase
the interest of the user. We noticed that the most famous accounts prefer the introductory strategy while the others
prefer a simple greeting. Quora members instead are used to a formal asking tone in contrast to \mbox{Twitter} where
people tend to converse in a more relaxed and friendly way. Figure~\ref{fig:Questions} shows an example of asking the
same question on topic \emph{motorbiking} on Twitter and Quora.

The question promotion process was easier in Quora because it is intentionally designed for Q\&A. We directly
pointed our tasks to the selected members through the built-in interface in Quora. Afterwards, if the question became
popular, it was further promoted by other members not related to our account or it became more visible because of the
high number of views and followers. Given this, we can conclude that the responsiveness of Quora is higher with respect
to Twitter which is also shown in the responsiveness rates of its users. Twitter on the other hand has a lower average
response rate due to its dynamic nature.

The main conclusion of this part of the work is that the networks that are primarily designed for task-solving need less
human intervention for both the bootstrapping and the promotion phases because many necessary steps like introduction,
thanking, rating, and rewarding are inherently present. Crowds of a more general purpose require additional human steps
in the workflow, otherwise people tend to be reluctant to help. Indeed, in earlier steps of this project when we did not
include any greeting, introductory or thanking messages the interaction was not satisfactory.

\subsection{Task Routing}

Table~\ref{tab:crowd_comparison} shows the main results from routing tasks to the targeted crowds. We received answers
to 44\% of the questions in Twitter and to 85\% of the questions in Quora. Nevertheless, only 64\% of the answers in
Quora came from the users we pointed. The rest were given by other users interested in the same topic. Similarly to what
we show in the Human vs. Machines section, the less popular questions were answered exclusively by only the users that
we targeted. This category constitutes 35\% of the whole questions posted. The answers' accuracy was manually evaluated
and varies between \mbox{80\%--90\%} for both of the networks which confirms that when people feel confident to answer
they are able to provide accurate insights.

Another major difference between the two crowds consists on the type of questions they can accommodate. Due to the
message length restrictions in Twitter, it is possible to ask only short questions that can be answered with short
replies. Such examples are queries looking for a fact or a single suggestion. This restriction does not make Twitter
users less expert. In fact, they also provided reasoning with their answers when they were able to explain in 140
characters. The answers in Quora are more elaborated and accordingly argued. In both networks the members preferred to
answer questions related to specific topics like \emph{biking}, \emph{hiking}, \emph{poker}, \emph{Bay Area} rather than
general ones like \emph{music}, \emph{sport}, \emph{travel}. A possible reason for this is that people tend to answer
more on topics in which they have experience and are particularly enthusiastic. This phenomenon constitutes an
important implicit incentive for most of the Q\&A applications and also for our study.

%% file: sections/sec8_conclusions.tex
\section{Conclusions and future work}
\label{sec:concl}

In this paper, we proposed a general model for task routing in online crowds that combines expertise detection with social availability
features. Furthermore, we presented the design and implementation of CrowdSTAR, a social task routing system. CrowdSTAR routes
questions to experts in the right crowd that are responsive in answering these questions. Yet, the system makes sure to
not overload experts with requests by regulating the number of questions routed to individual users. CrowdSTAR currently
supports two popular social networks, Twitter and Quora, but the architecture is extensible to other crowds.

Our findings show that the proposed user utility model exists in real social networks and can be used to increase answer
satisfaction during question routing.  The evaluation shows that experts are willing to answer questions which are more
specific. As future work, we plan to investigate reward mechanisms to give experts further incentives to answer questions and improve the
crowd selection algorithm using adaptive learning techniques. Moreover, we want to investigate the problem of evaluating
accuracy of task routing and how intrinsic features of social networks can be used in this respect.